\documentclass[12pt]{article}

\usepackage{latexsym,amsmath,amssymb,theorem,epsfig} 

\DeclareFontFamily{OT1}{rsfs10}{} 
\DeclareFontShape{OT1}{rsfs10}{m}{n}{ <-> rsfs10 }{} 
\DeclareMathAlphabet{\mathscript}{OT1}{rsfs10}{m}{n} 

\numberwithin{equation}{section}

\newcommand{\ns}{\normalsize}

\newcommand{\tr}{\text{tr}}

\def\b{\beta}

\def\e{\epsilon}

\theoremstyle{plain} 

{\theorembodyfont{\rmfamily} }

\begin{document}


\begin{titlepage}

\vspace{-5cm}

\title{
  \hfill{\ns }  \\[1em]
   {\LARGE Vacuum Stability in Heterotic M-Theory} 
\\[1em] } 
\author{
   Evgeny I. Buchbinder$^{1}$ 
   and Burt A. Ovrut$^{2}$  \\[0.5em]
   {\ns $^1$School of Natural Sciences, Institute for Advanced Study} \\[-0.4cm]
{\ns Einstein Drive, Princeton, NJ 08540}\\[0.3cm]
{\ns $^2$Department of Physics and Astronomy, University of 
Pennsylvania} \\[-0.4em]
   {\ns Philadelphia, PA 19104}\\}
      
\date{}

\maketitle

\begin{abstract}

The problem of the stabilization of moduli is discussed within the context 
of compactified strongly coupled heterotic string theory. It is shown 
that all geometric, vector bundle and five-brane moduli are completely fixed, 
within a phenomenologically acceptable range, by non-perturbative physics. 
This result requires, in addition to the full space of moduli, non-vanishing 
Neveu-Schwarz flux, gaugino condensation with threshold corrections 
and the explicit form of the Pfaffians in string instanton superpotentials. 
The stable vacuum presented here has a negative cosmological constant. 
The possibility of ``lifting'' this to a metastable vacuum with positive 
cosmological constant is briefly discussed.
 
\end{abstract}

\thispagestyle{empty}

\end{titlepage}


\section{Introduction}


One of the most important problems in finding realistic
four-dimensional vacua in superstrings and M-theory is the problem 
of moduli stabilization. The four-dimensional
fundamental constants, such as the Newton and unification
gauge parameters, depend on the compactification moduli. Therefore,
in any  realistic compactification scenario, all the moduli 
have to be fixed, or very slowly rolling, in a 
phenomenologically acceptable range. However, string theory 
moduli do not have a perturbative potential energy. Hence, 
if their values are to be fixed, it must be by non-perturbative physics.
The first attempts to do this~\cite{DSWW1, DSWW2}
indicated that non-perturbative superpotentials 
can lead to runaway behaviour. 
That is, the radius of the compactification manifold was found to 
run to large values, leading to decompactification. However, 
this work was very preliminary, involving only a subset of possible 
moduli and non-perturbative superpotentials.

Over years, there have been many attempts to prove the stability 
of moduli in different types of string theory. Recently,
progress in this direction was achieved in type IIB string theory 
in~\cite{Kachru}, emphasizing, among other things, 
the necessity of considering flux 
compactifications ~\cite{Sethi1, Gid, Verlinde, Schulz, Schulz2, 
Frey, GVW, Bobby, Taylor, Curio, Cardoso, Nonkahler1, Nonkahler2, Nonkahler3, Curiok1, Curiok2}. 
The moduli stabilization in~\cite{Kachru}
was demonstrated in two steps. First, all moduli were stabilized at a fixed 
minimum with a negative cosmological constant. This was achieved 
by combining fluxes with non-perturbative effects. Second, the minimum was 
lifted to a metastable vacuum with a positive cosmological 
constant. This was  accomplished by adding anti D-branes and using 
previous results, obtained in~\cite{Verlinde}, that the flux-anti D-brane 
system can form a metastable bound state with positive energy. 
In ref.~\cite{Kachru}, it was also shown that one can fine tune
various parameters to make the value of the cosmological constant
consistent with the observed amount of dark energy. 

In this paper, we consider the problem of moduli stabilization in 
strongly coupled heterotic string theory~\cite{HW1, HW2} 
compactified on Calabi-Yau threefolds.
Such compactifications are called heterotic M-theory and
have a number of phenomenologically attractive 
features (see~\cite{Far} for a recent review of the  
phenomenological aspects of M-theory).
In~\cite{DOPW1, DOPW2, DOPW3, DOPW4}, a specific 
set of vacua were constructed consisting of appropriate $SU(5)$ vector 
bundles over Calabi-Yau threefolds with 
${\mathbb Z}_2$ fundamental group. 
These lead 
to four-dimensional theories with the standard model
$SU(3)\times SU(2) \times U(1)$
gauge group and three families of charged chiral matter.
Recently, in refs.~\cite{Rene1, Rene2, Rene3}, 
these theories were generalized to vacua involving $SU(4)$ bundles over
Calabi-Yau threefolds with  
${\mathbb Z}_2 \times {\mathbb Z}_2$ fundamental group. Such vacua 
correspond to standard model-like physics in four dimensions with 
potentially suppressed nucleon decay. In this paper, for simplicity,
we consider only vector bundles over simply connected Calabi-Yau 
threefolds, compactifications which are easier to analyze. However, 
we see no reason why our results should not apply to more realistic 
heterotic vacua on non-simply connected manifolds. Within this context, 
we will consider all geometric and vector bundle moduli. In addition, we include the translational 
moduli of M five-branes.
One of the features of strongly coupled heterotic 
string compactifications is the presence of five-branes.
In~\cite{Nonstandard, Lalak}, it was argued that vacua 
with five-branes are more natural since, for example, it is much 
easier to satisfy the anomaly cancellation 
condition in their presence. Since, in order to obtain
phenomenologically acceptable values for the fundamental constants one 
has to take
the size of the eleventh-dimension to be larger than the Calabi-Yau 
scale~\cite{Wittenstrong, Banks}, the translational modes of the five-branes 
will appear as moduli in the four-dimensional low-energy effective action.

In this paper, we show that all 
heterotic M-theory moduli, that is, the complex structure, Kahler, 
vector bundle and five-brane moduli, can be 
stabilized by non-perturbative superpotentials.
Recent discussions of this issue~\cite{Moore}, in models with a restricted
number of moduli, indicated instabilities caused by membrane 
instantons~\cite{BBS, Witten96}. 
However, these models did not include all
compactification and vector bundle moduli, as well as all possible 
sources for superpotentials.
The analysis of~\cite{Moore} was refined in~\cite{Curio1}, where it was shown 
that stabilization of certain moduli can be achieved. Nevertheless, 
again, not all moduli were taken into account. In addition, 
the authors in~\cite{Curio1} 
chose various parameters outside their natural range.
In the present paper, we show that including all moduli
and all superpotentials does lead to complete moduli stabilization in 
a phenomenologically accepted range with a negative cosmological constant. 
Furthermore, this is achieved within the natural values of the parameters.
Our stabilization procedure uses various tools such as form 
flux, non-perturbative superpotentials including their 
bundle moduli dependent Pfaffians and gaugino condensation
on the hidden brane with threshold corrections. Even though supersymmetry
is not broken in the moduli sector, it is softly broken in the 
gravity and matter sector at the TeV scale by the gaugino 
condensate~ \cite{Nonstandard, Dine, KL, BIM, NOY, LT, Horava, LOW1, LOW2}.

The paper is organized as follows. In Section 2, we review 
the Kahler potentials for all heterotic string moduli with a
detailed discussion of their relative scales, including 
the vector bundle moduli Kahler potential. 
Even though the latter cannot be evaluated explicitly for 
vector bundles on a threefold\footnote{In fact, in ref.~\cite{Grey},
the vector bundle moduli Kahler potential 
was approximately computed for special types of Calabi-Yau threefolds and 
very special types of bundles. The bundles considered in~\cite{Grey} 
were taken to be the pullback of vector bundles on a surface. Such
bundles admit a gauge connection which is approximately ADHM, provided
the instanton is sufficiently small.
To a generic bundle on a threefold which does not come from a bundle
on lower dimensional space, the method of~\cite{Grey} cannot be applied.},
it is possible to derive the relevant properties we will need in 
later sections. 
In Section 3, we give a careful analysis of the superpotentials with
a detailed discussion of their scales. In Subsection 3.1, we discuss
the flux-induced superpotential derived in~\cite{Behrndt, Constantin}.
This superpotential depends on the complex structure moduli. 
In Subsection 3.2, we introduce the superpotential induced 
by a gaugino condensate on the hidden 
brane~\cite{Nonstandard, Dine, KL, BIM, NOY, LT, LOW1, LOW2}. 
This superpotential depends on the Calabi-Yau
volume as well as on the size of the eleventh dimension and 
the five-brane moduli through the threshold corrections. In Subsection 3.3, 
we discuss various non-perturbative superpotentials induced by 
membrane instantons~\cite{BBS, Witten96}. They 
depend on the $(1,1)$ moduli, the five-brane 
moduli and the vector bundle moduli. Various pieces of these
superpotentials were calculated in~\cite{Moore, Lima1, Lima2, BDO1, BDO2}. 
In order to obtain the total superpotential, one 
has to sum over contributions coming from all genus zero holomorphic curves in 
a given Calabi-Yau threefold.
We give arguments that, in the models under consideration, the 
superpotential does not vanish after summation. 
In Section 4, we show, in detail, that all heterotic M-theory 
moduli can be fixed at 
a stable AdS minimum. In Subsection 4.1, we set up the model.
Since the Kahler potentials and the superpotentials are 
very complicated when the number of $(1,1)$ and vector bundle 
moduli is large, we have to introduce some
simplification in order to obtain an analytic solution. 
We argue that if we restrict ourselves to consider 
only one $(1,1)$ modulus, which coincides with the size of the 
eleventh dimension, and make some further restrictions on the number 
of the vector bundle moduli, we do not actually lose generality. 
Finally, in Subsection 4.2, we prove the stabilization of the moduli. 
The complex structure moduli and the Calabi-Yau volume are 
stabilized by a mechanism similar to that considered in~\cite{Bobby} and~\cite{Kachru}. 
In addition, we show that the vector bundle and five-brane moduli are also 
fixed. We then analyze how our equations would be modified 
if we had many $(1,1)$ and vector bundle moduli. We conclude that we
would still find a stable solution and, hence, that the restriction 
to a single $(1,1)$ modulus and one vector bundle modulus was without 
loss of generality. 
There is only one mild constraint that we have to impose 
on a single coefficient to make sure that the five-brane modulus 
is stabilized in an acceptable range. In the Conlusion, we summarize 
our results. We also discuss the possibility 
of lifting our minimum to a metastable vacuum with a positive 
cosmological constant, as was done in~\cite{Kachru}. These 
results will apear elsewhere.


\section{The Kahler Potentials}


We consider $E_8 \times E_8$ strongly coupled heterotic superstring 
theory~\cite{HW1, HW2} on the space 
\begin{equation}
M={\mathbb R}^4 \times X \times S^{1}/Z_2,
\label{1.1}
\end{equation}
where $X$ is a Calabi-Yau threefold. Let us list
the complex moduli fields arising from such a compactification. They are 
the $h^{1,1}$ moduli $T^{I}$, the volume modulus $S$, the 
$h^{1,2}$  moduli $Z_{\alpha}$ and the vector bundle moduli
which we denote by $\Phi_{u}$. 
In addition, we will assume that anomaly cancellation 
requires the existence of a non-trivial five-brane class. 
Furthermore, for simplicity, we will work in the region of its moduli space 
corresponding to a single five-brane~\cite{Nonstandard, Five}. The five-brane translational
complex modulus will be denoted by
${\bf Y}$. 
In this section, we review 
the Kahler potentials  for the $T^{I}, S, Z_{\alpha}$ and ${\bf Y}$ moduli and 
derive some general properties of the vector bundle moduli Kahler 
potential. 

The moduli $T^I$ are defined as 
%
\begin{equation}
T^{I} = Ra^{I}V^{-1/3}+ \frac{{\rm i}}{6} p^{I},
\label{1.2}
\end{equation}
where $R$ is the orbifold plane separation modulus, 
$V$ is the Calabi-Yau 
breathing modulus, $a^{I}$ are the $(1,1)$ moduli of the Calabi-Yau space 
and the imaginary parts $p^I$ arise from the eleventh component 
of the graviphotons. 
The Calabi-Yau breathing modulus $V$ also appears as the real part 
of the four-dimensional dilaton multiplet 
\begin{equation}
S=V+{\rm i}  \sqrt{2} \sigma,
\label{1.3}
\end{equation}
where the imaginary part $\sigma$
originates from dualizing the four-dimensional 
$B$-field. The moduli $a^I$ and $V$ are not independent. It can be shown 
that 
\begin{equation}
V=\frac{1}{6} \sum_{I,J,K=1}^{h^{1,1}}d_{IJK}a^I a^J a^K,
\label{1.4}
\end{equation}
where $d_{IJK}$ are the intersection numbers of the Calabi-Yau threefold.
Note that the moduli $V$ and $R$ are dimensionless and defined as 
\begin{equation}
V=\frac{1}{v_{CY}}\int_{CY} \sqrt{g_{CY}}
\label{1.5}
\end{equation}
and 
\begin{equation}
R=\frac{1}{\pi \rho}\int dx^{11}
\label{1.6}
\end{equation}
respectively.
Here $v_{CY}$ is the reference volume of the Calabi-Yau threefold, 
$\pi \rho$ is the reference length of the eleventh dimension and $x^{11}$ is the coordinate
along the interval $S^1/Z_2$.
The actual volume of the threefold and the actual size of the 
eleventh dimension are $v_{CY}V$ and $\pi \rho R$ respectively.
See~\cite{Univ, LOSW5, LOSW4}
for more details on the compactification of strongly coupled
heterotic string theory to five and 
four dimensions and the structure of the chiral multiplets.
To achieve the correct phenomenological values for the 
four-dimensional Newton and gauge coupling parameters,
\begin{equation}
M_{Pl} \sim 10^{19}GeV, \quad \alpha_{GUT} \sim \frac{1}{25}, 
\label{check}
\end{equation}
we assume~\cite{Wittenstrong, Banks} 
that the inverse reference radius of the Calabi-Yau threefold
and the inverse reference length 
of the eleventh dimension are
\begin{equation}
v_{CY}^{-1/6} \sim 10^{16} GeV, \quad (\pi \rho)^{-1} \sim 10^{14} GeV
\label{doublecheck}
\end{equation}
respectively. This implies 
that, at the present time, 
the dimensionless moduli $V$ and $R$ have to 
be stabilized at, or be very slowly rolling near,
the values
%
\begin{equation}
V \sim 1, \quad T \sim 1.
\label{values}
\end{equation}

The Kahler potential for $S$ and $T^{I}$ moduli was computed 
in~\cite{LOSW4}. It is given by 
\begin{equation}
K_{S, T} = -M^{2}_{Pl} \ln (S+\bar S) - 
M^{2}_{Pl} \ln \left (\frac{1}{6} \sum_{I,J,K =1}^{h^{1,1}}
d_{IJK}(T+ \bar T)^{I}(T+ \bar T)^{J}  (T+ \bar T)^{K} \right ).
\label{1.7}
\end{equation}
The Kahler potential for the complex structure moduli $Z_{\alpha}$ 
was found in~\cite{Candelas} to be 
\begin{equation}
K_{Z}= -M^{2}_{Pl} \ln (-i \int_{X} \Omega \wedge \bar \Omega ),
\label{1.8}
\end{equation}
where $\Omega$ is the holomorphic (3,0) form.

The next supermultiplet to discuss is the one associated with the five-brane  
modulus ${\bf Y}$. It was shown in~\cite{Nonstandard} that, when a five-brane is 
compactified to four-dimensions on a holomorphic curve $z$ of genus $g$, there are
two types of zero-mode supermultiplets that arise. First, there are $g$ Abelian 
vector superfields which are not of our interest in this paper. The second multiplet that arises 
is associated with the translational scalar mode $Y$. Geometrically
$Y$ corresponds to the position of the five-brane in the fifth 
dimension\footnote{Note that since $v_{CY}^{1/6} \ll \pi \rho$, the eleventh 
dimensional coordinate $x^{11}$ parametrizes the fifth dimension of the effective 
five-dimensional theory.}.
It was shown in~\cite{Deren} that the translational multiplet of the 
five-brane is a chiral supermultiplet whose bosonic component ${\bf Y}$ is given by
\begin{equation}
{\bf Y}=\frac{Y}{\pi \rho}Re {\cal T} +i (a + \frac{Y}{\pi \rho} Im {\cal T}).
\label{1.8.1}
\end{equation}
Here $a$ is the axion arising from dualizing the three-form field strength propagating
on the five-brane world-volume and 
${\cal T}$ is related to the $(1,1)$-moduli $T^{I}$ as follows. 
Let $\omega_{I}, I=1, \dots h^{1,1}$ be a basis of harmonic $(1,1)$ forms 
on our  Calabi-Yau threefold. These are naturally dual to a basis $z^{I}, I=1, \dots h^{1,1}$ 
of curves in $H_{(1,1)}(X)$ where
\begin{equation}
\frac{1}{v_5} \int_{z^{I}} \omega_{J} =\delta_{IJ}.
\label{1.8.2}
\end{equation}
Parameter $v_5$ is the volume of the curve $z$
on which the five-brane is wrapped. Any holomorphic curve can be expressed as 
a linear combination of the $z^{I}$ curves. The curve on which the five-brane is 
wrapped can be written as 
\begin{equation}
z =\sum_{I=1}^{h^{(1,1)}} c_I z^I
\label{1.8.3}
\end{equation}
for some coefficients $c_I$. 
The modulus ${\cal T}$ which appears in~\eqref{1.8.1} is defined as 
\begin{equation}
{\cal T} = \sum_{I=1}^{h^{(1,1)}} c_I T^I.
\label{1.8.4}
\end{equation}
The Kahler potential for the ${\bf Y}$ modulus was calculated in~\cite{Deren}
and found to be 
\begin{equation}
K_5 = 2 M^{2}_{Pl} \tau_5 \frac{({\bf Y} +\bar{\bf Y})^2}{(S+\bar S)({\cal T}+\bar {\cal T})},
\label{1.8.5}
\end{equation}
with the coefficient $\tau_5$ given by 
\begin{equation}
\tau_5 =\frac{T_5 v_5 (\pi \rho)^2}{M^{2}_{Pl}}
\label{1.8.6}
\end{equation}
and $T_5$ is 
\begin{equation}
T_5 =(2 \pi)^{1/3} (\frac{1}{2 \kappa_{11}^2})^{2/3}, 
\label{1.8.7}
\end{equation}
where $\kappa_{11}$ is the eleven-dimensional gravitational coupling constant. 
It is related to the four-dimensional Planck 
mass as 
\begin{equation}
\kappa_{11}^2=\frac{\pi \rho v_{CY}}{M^2_{Pl}}.
\label{1.15}
\end{equation}
If we substitute eq.~\eqref{1.15} into~\eqref{1.8.7}  
using~\eqref{check} and~\eqref{doublecheck},
%
%
we obtain 
\begin{equation}
\tau_5 \approx \pi \frac{v_5}{v_{CY}^{1/3}}.
\label{1.8.8}
\end{equation}

Now let us move on to the vector bundle moduli Kahler potential. Its general expression
can be 
obtained from the dimensional reduction of the term in the ten-dimensional action 
\begin{equation}
\frac{-1}{4 g_{10}^{2}}{\rm tr}
\int d^{10}x \sqrt{-g} F_{MN}^{2},
\label{1.9}
\end{equation}
where $M, N = 0, 1, \dots , 9$ and  
$g_{MN}$ and $F_{MN}$ are the ten-dimensional metric and 
Yang-Mills field strength respectively.
Upon dimensional reduction, the ten-dimensional 
metric and gauge field split as follows 
\begin{eqnarray}
&& d s^{2}_{10} =
v_{CY}^{-\frac{2}{3}}
g_{\mu \nu} dx^{\mu}dx^{\nu} + v_{CY}^{\frac{1}{3}}
g_{CY m \bar m}dX^{m}  
d \bar X^{\bar m},
\nonumber \\
&& A_{M} = (A_{\mu}, A_{m}, \bar A_{\bar m}),
\label{1.10}
\end{eqnarray}
where $\mu, \nu = 0, 1, 2, 3$ and $m, \bar m =1, 2, 3$. The fields 
$g_{\mu \nu}$ and $A_{\mu}$ are the 
four-dimensional metric and the gauge field respectively,
whereas $g_{m \bar m}$ and $A_{m}$ represent the metric and the gauge 
connection on the Calabi-Yau threefold. 
Substituting~\eqref{1.10} into the action~\eqref{1.9},
we obtain the following expression for the vector bundle moduli Kahler 
potential
\begin{equation}
\tilde{K}_{bundle} =
\frac{1}{2g_{10}^2} {\tr} \int d^{6}X
\sqrt{g_{CY}} g^{m \bar m} A_{m} \bar A_{\bar m}.
\label{1.11}
\end{equation}
Let us find the scale that controls the strength of the 
Kahler potential. To do this, introduce 
the dimensionless quantities 
\begin{equation}
\tilde{X}^{m} =\frac{X^{m}}{v_{CY}^{1/6}}, \quad 
\tilde{A}_{m} =A_{m}v_{CY}^{1/6},
\label{1.12}
\end{equation}
where $v_{CY}$ is the Calabi-Yau reference volume.
We also normalize all vector bundle moduli associated with
$A_{m}$ with respect to the Calabi-Yau reference volume so that
they too are dimensionless.
The Kahler potential then becomes 
\begin{equation}
\tilde{K}_{bundle}=
\frac{v_{CY}^{2/3}}{2g_{10}^2} {\tr} \int d^{6}\tilde{X}
\sqrt{g_{CY}} g^{m \bar m} 
\tilde{A}_{m} 
\stackrel{-}{\tilde{A}_{\bar m}}.
\label{1.13}
\end{equation}
The ten-dimensional gauge coupling parameter is related to the 
eleven-dimensional Planck scale as~\cite{HW1, HW2}
\begin{equation}
\frac{1}{g^{2}_{10}} = \frac{1}{2\pi \kappa_{11}^2}
(\frac{\kappa_{11}^2}{4\pi})^{2/3}.
\label{1.14}
\end{equation}
%
%
From equations~\eqref{1.15} and~\eqref{1.14} we obtain 
\begin{equation}
\frac{v_{CY}^{2/3}}{2g_{10}^2} =
\frac{1}{(4\pi)^{5/3}}
\frac{M_{Pl}^2}{((\pi \rho)^2 M^2_{Pl})^{1/3}}.
\label{1.16}
\end{equation}
We can then write the Kahler potential $\tilde{K}_{bundle}$ as 
\begin{equation}
\tilde{K}_{bundle}=k M^{2}_{Pl} K_{bundle},
\label{1.17}
\end{equation}
where 
\begin{equation}
k=\frac{1}{(4\pi)^{5/3} 
((\pi \rho)^2 M^2_{Pl})^{1/3}}
\label{1.18}
\end{equation}
and 
\begin{equation}
K_{bundle}=
 {\tr} \int d^{6}\tilde{X}
\sqrt{g_{CY}} g^{m \bar m} \tilde{A}_{m} 
\stackrel{-}{\tilde{A}_{\bar m}}.
\label{1.19}
\end{equation}
Note that $K_{bundle}$ is dimensionless since it depends on dimensionless
vector bundle moduli. The parameter $k$ is also dimensionless.
Substituting~\eqref{check} and~\eqref{doublecheck} into~\eqref{1.18}, 
we obtain 
\begin{equation}
k \sim 10^{-5}.
\label{1.20}
\end{equation}

The reason that the strength of the vector bundle moduli 
Kahler potential is smaller by several orders of magnitude than 
the strength of the $T, S$ and $Z$ Kahler potentials is that 
the $F^2$ term appears to the next order in $\alpha^{\prime}$ 
in the ten-dimensional action as comparing to the supergravity multiplet.
Unfortunately, to the same order in $\alpha^{\prime}$ in the 
ten-dimensional action and, as a consequence, to the same order
in $k$ in the four-dimensional action, there is a cross term between the 
$T$-moduli and the vector bundle moduli.
This cross term comes from the 
\begin{equation}
\int d^{10} x \sqrt{-g} H \wedge H^{*}
\label{cross1}
\end{equation}
term in the ten-dimensional action,
where $H$ is given by~\cite{HW2} (see also~\cite{LOSW4})
\begin{equation}
H = dB-\frac{1}{4 \sqrt{2} \pi^2 \rho} 
(\frac{\kappa_{11}}{4 \pi})^{2/3}
(\omega_{YM} - \omega_{L})
\label{cross2} 
\end{equation}
and $\omega_{YM}$ and $\omega_{L}$ are the Yang-Mills and gravitational Chern-Simons
forms respectively. The term~\eqref{cross1} leads to the following 
contribution to the four-dimensional effective action,
\begin{equation}
\sim k M^2_{Pl}\int d^{4}x \sqrt{-g_4} g^{\mu \nu} \sum_{I=1}^{h^{1,1}} (Im T^{I} )
\int d^6 \tilde{X} \sqrt{g_{CY}}  \omega_{I}^{ m \bar m} \partial_{\mu} 
\tilde{A}_{m} \partial_{\nu}
\stackrel{-}{\tilde{A}_{\bar m}}.
\label{cross3}
\end{equation}
In this expression, $\tilde{X}$ and $\tilde{A}$ are the rescaled 
Calabi-Yau coordinates and gauge connection~\eqref{1.12},
$\omega_{I m \bar m}$ are the basis of the harmonic $(1,1)$ forms on the Calabi-Yau threefold 
and the coefficient $k$ is given precisely by~\eqref{1.18}.
This cross term does not significantly effect the Kahler potential 
for the $T$-moduli, since it appears at a lower scale. However, 
it does effect the vector bundle moduli Kahler potential~\eqref{1.17}-\eqref{1.19}. 
Schematically,  the pure vector bundle moduli Kahler metric can be written as
\begin{equation}
\int_{X} \partial \tilde{A} \bar \partial \stackrel{-}{\tilde{A}},
\label{cross4}
\end{equation}
whereas
the cross term can be written as
\begin{equation}
\sum_I (ImT^I) \int_{X} \omega_{I} \partial \tilde{A} \bar \partial \stackrel{-}{\tilde{A}}.
\label{cross5}
\end{equation}
It is clear that the cross term can be ignored as long as the values of the imaginary parts 
of the $T$-moduli are sufficiently smaller than one. For now, we will simply assume
that this is the case and discard the cross term. Later, when studying stabilization issues, we will
see that one can indeed stabilize the imaginary parts of the $T$-moduli
at values sufficiently less than one, thus justifying our assumption.

It is difficult to calculate the vector bundle moduli part of the Kahler potential
explicitly without knowing a solution to the hermitian 
Yang-Mills equations. 
Nevertheless, some properties of
$K_{bundle}$ can be determined. These properties will be 
sufficient to allow one to
study the issues of moduli stabilization in later sections.
At this point, we have to be more specific about the type of Calabi-Yau threefold  
we choose and the type of vector bundle we put over it. 
In this paper, the Calabi-Yau threefold will be taken to be elliptically fibered.
For such Calabi-Yau spaces, there exists a rather explicit spectral cover construction
of stable holomorphic vector bundles~\cite{FMW, Donagi}. The moduli of such 
vector bundles were discussed in~\cite{BDOold}. 
In the present paper, we will restrict our discussion to such vector bundles.
Geometrically, their moduli space is just
a complex projective space ${\mathbb C}{\mathbb P}^{N}$, where 
$N$ is the total number of the vector bundle moduli~\cite{BDOold}. 
The moduli of vector bundles on elliptically fibered Calabi-Yau manifolds
will be reviewed in more detail in the next section. For now, we will 
only need the fact that the moduli parametrize a complex projective space.
Strictly speaking, 
the moduli space of bundles ${\cal M}$ is an open subset in 
${\mathbb C}{\mathbb P}^{N}$. The projective space is actually 
the compactification of ${\cal M}$ with respect to certain singular 
objects known as torsion free sheaves.
The gauge connection becomes singular on these sheaves.
However, for simplicity, we will view
${\mathbb C}{\mathbb P}^{N}$ as the moduli space of vector bundles, 
keeping in mind that it also contains singular points. 
At these points, 
the Kahler potential should blow up since 
the associated gauge connections do.
As some of the vector bundle moduli approach certain critical values,
the corresponding gauge connection represents a delta-function peak over some holomorphic 
curve in the Calabi-Yau threefold. 
These moduli are called the transition moduli associated with this curve~\cite{BDOold}.
We will cover our  ${\mathbb C}{\mathbb P}^{N}$ manifold
with standard open sets isomorphic to ${\mathbb C}^{N}$
by introducing $N+1$ homogeneous 
coordinates and setting one of them to unity on one of the open sets.
Let us consider any open patch $U_{\alpha} \subset {\mathbb C}{\mathbb P}^N$
containing 
the transition moduli associated with some holomorphic
curve. Denote this curve by 
$z$ 
and let the number of the transition moduli be $M$. 
Let the $N$ local coordinates on this open 
set be $\Phi_u=(\phi_{i}, \psi_a)$, where $\phi_i$ represent the transition
moduli of the curve $z$ and $\psi_a$ the remaining moduli.
The total number of parameters is, of course, $N$. 
One can always choose the coordinate system in such a way that the critical values 
of the transition moduli are 
\begin{equation}
\phi_i =0, \quad i=1, \dots , M.
\label{1.21}
\end{equation}
The codimension 
$N-M$ subset of ${\mathbb C} {\mathbb P}^N$ defined by these equations
%
%
represents a singularity of the type described above. 
When all of the $\phi_i$ go to zero,
the bundle becomes a singular torsion free sheaf. 
This corresponds 
to the gauge connection being a distribution, that is infinitely peaked about the 
$z$ curve and 
smooth everywhere else. As one turns the moduli 
$\phi_i$ on, the torsion free sheaf smears out to produce a smooth vector
bundle with an everywhere smooth hermitian connection. 
It is clear that at the torsion free sheaf, where
the gauge connection has an infinite peak centered 
at $z$, the Kahler potential~\eqref{1.19}
diverges. Note that this is generically true at any 
singular point in moduli space.
The above analysis allows us to say that for values of $\phi_i$ sufficiently small, 
we can approximately split the Kahler potential $K_{bundle}$ as
\begin{equation}
K_{bundle} = K_{bundle}(\phi) + K_{bundle}(\psi).
\label{1.21.1}
\end{equation}
The reason is that for $\phi_i$'s small enough, the gauge connection can approximately
be written as 
\begin{equation}
A = A(\phi) + A(\psi),
\label{1.22.2}
\end{equation}
where $A(\phi)$ is strongly centered around the curve $z$ and $A(\psi)$ is smooth everywhere. 
In the limit of small $\phi_i$, the overlap integral of the product
of these two pieces of the gauge connection
is small. Then~\eqref{1.21.1} follows from~\eqref{1.22.2} and~\eqref{1.19}.
We will also need to know what happens to $K_{bundle}$ as the moduli 
(either $\phi_i$ or $\psi_a$) become large in the sense of coordinates
on ${\mathbb C}^{N}$. Since 
\begin{equation}
h^{1,1}({\mathbb C}{\mathbb P}^{N})=1,
\label{1.22}
\end{equation}
there exists a unique cohomology class of Kahler forms.
The Kahler form associated with the well-known Fubini-Study Kahler metric on
${\mathbb C}{\mathbb P}^{N}$ is contained in this 
non-trivial cohomology class. This 
means that every Kahler potential on ${\mathbb C}{\mathbb P}^{N}$ 
can be written as 
\begin{equation}
K_{{\mathbb C}{\mathbb P}^{N}}=K_{FS}+f,
\label{1.23}
\end{equation}
where $K_{FS}$ is the Fubini-Study Kahler potential and $f$ is any 
global function. The only restriction on $f$ is that the corresponding 
Kahler metric has to be positive definite. On the coordinate patch 
$U_{\alpha}$ with local coordinates $\Phi_{u}$, we have
\begin{equation}
K_{{\mathbb C}{\mathbb P}^{N}}|_{U_{\alpha}}= 
K_{FS}|_{ U_{\alpha}}+ f \rho_{\alpha},
\label{1.24}
\end{equation}
where
\begin{equation}
K_{FS}|_{U_{\alpha}} = \ln (1+ \sum_{u=1}^{N}|\Phi_{u}|^2)
\label{1.25}
\end{equation}
and $\{\rho_{\alpha}\}$ 
is the partition of unity. As we approach the boundary 
of the open set, 
\begin{equation}
\rho_{\alpha} \rightarrow 0.
\label{1.26}
\end{equation}
Furthermore, from~\eqref{1.25}
we find that 
\begin{equation}
K_{{\mathbb C}{\mathbb P}^{N}}|_{U_{\alpha}}\rightarrow \infty
\label{1.27}
\end{equation}
in this limit.
From this analysis, we conclude that $K_{bundle}$ grows as one increases
either one of the $\phi_i$'s or one of the $\psi_a$'s keeping the other 
variables fixed.
These properties of $K_{bundle}$ will be important in the next sections.


\section{Superpotentials}


In this section, we discuss the superpotentials that will be used 
to achieve the stabilization of all moduli considered above. 


\subsection{The Flux-Induced Superpotential}


We want to turn on a non-zero flux of the Neveu-Schwarz three-form 
$H$ on the Calabi-Yau threefold. The presence of this non-zero flux 
generates a superpotential for the $h^{1,2}$ moduli of the form~\cite{Behrndt, Constantin}
\begin{equation}
W_{f} \sim  \int_{X} H \wedge \Omega.
\label{2.1}
\end{equation}
This is the heterotic analog of the type IIB superpotential~\cite{GVW, Taylor, Curio}
\begin{equation}
W_{IIB} \sim \int_{X} G_3 \wedge \Omega,
\label{2.2}
\end{equation}
where $G_3 = F_3 - \tau H_3$. 
Expression~\eqref{2.1} can be obtained by considering the variation of the 
ten-dimensional gravitino, 
dimensionally reducing this to four-dimensions and 
matching it against the well-known 
gravitino transformation law in four-dimensional supergravity. 
See ref.~\cite{Constantin} for a detailed derivation.

For later use, we need to find the scale that controls $W_f$. Since 
the components of $H$ have dimension one, we find that 
\begin{equation}
W_{f} = \frac{M^2_{Pl}}{v_{CY}} \int_{X} H \wedge \Omega.
\label{2.3}
\end{equation}
As before, introduce dimensionless coordinates $\tilde{X}^m$ 
\begin{equation}
\tilde{X}^m=\frac{X^m}{v_{CY}^{1/6}}
\label{2.4}
\end{equation}
and dimensionless components for the three-form. 
Since $H$ is quantized in units of~\cite{Wittenflux} 
\begin{equation}
(\frac{\kappa_{11}}{4\pi})^{2/3} \frac{1}{\pi \rho},
\label{quanta}
\end{equation}
the components 
of $H$ and the dimensionless three-form $\tilde{H}$ are related by
\begin{equation}
H_{mnp} =(\frac{\kappa_{11}}{4\pi})^{2/3} \frac{1}{\pi \rho v_{CY}^{1/2}}\tilde{H}_{mnp}.
\label{2.5}
\end{equation}
As a consequence, $W_{f}$ can be written as 
\begin{equation}
W_{f} = \frac{M^2_{Pl}}{v_{CY}^{1/2}} 
(\frac{\kappa_{11}}{4\pi})^{2/3} \frac{1}{\pi \rho}
\int_{X} \tilde{H} \wedge 
\tilde{\Omega}= M^{3}_{Pl} h_1 \int_{X} \tilde{H} \wedge 
\tilde{\Omega},
\label{2.6}
\end{equation}
where, using eqs.~\eqref{check} and~\eqref{doublecheck}, we find
\begin{equation}
h_1 = \frac{1}{M_{Pl}^3 v_{CY}^{1/2}} \sim 2 \cdot 10^{-8}
\label{2.7}
\end{equation}
and $\tilde{H}$ and $\tilde{\Omega}$ are both dimensionless.

Note that turning on a non-vanishing flux warps the compactification space
away from a pure Calabi-Yau threefold. The strength of this warping is 
determined by the dimensionless parameter 
\begin{equation}
(\frac{\kappa_{11}}{4\pi})^{2/3} \frac{1}{\pi \rho v_{CY}^{1/2}}
\int_{C} \tilde{H},
\label{star}
\end{equation}
where $C$ is an appropriate three cycle. Since 
$(\frac{\kappa_{11}}{4\pi})^{2/3} \frac{1}{\pi \rho v_{CY}^{1/2}} \sim 2 \cdot 10^{-5}$, 
it follows that for 
\begin{equation}
\int_{C} \tilde{H} \ll \frac{1}{2} 10^5
\label{starstar}
\end{equation}
the warping away from a Calabi-Yau threefold is negligably small. 
Henceforth, we will always choose the flux to satisfy condition~\eqref{starstar}. 


\subsection{Gaugino Condensation Induced Superpotential}


We also turn on a gaugino condensate on the hidden 
brane
\cite{Nonstandard, Dine, KL, BIM, NOY, LT, LOW1, LOW2}. 
A non-vanishing
gaugino condensate has important phenomenological consequences.
Among other things, it is responsible for 
supersymmetry breaking in the hidden sector. When that symmetry 
breaking is transported to the observable brane, it leads to 
soft supersymmetry breaking terms
for the
gravitino, gaugino and matter fields on the order of the
electroweak scale.
A gaugino condensate is also 
relevant to the discussion in this paper, since it
produces a superpotential for $S$, $T$ and ${\bf Y}$ moduli of the form
\begin{equation}
W_g =M_{pl}^3 h_2 exp (-\e S +\e \alpha^{(2)}_{I} T^{I} -
\e \b \frac{{\bf Y}^2}{{\cal T}}).
\label{2.8}
\end{equation}
Here~\cite{LOW1} 
\begin{equation}
h_2 \sim \frac{1}{M_{Pl} \sqrt{v_{CY}}(\pi \rho)} (\frac{\kappa_{11}}{4 \pi})^{2/3}
\sim 10^{-6},
\label{2.9}
\end{equation}
and the coefficient $\e$ is related to the coefficient $b$ of
the one-loop beta function
and is given by 
\begin{equation}
\e = \frac{6 \pi}{b_0 \alpha_{GUT}}.
\label{2.10}
\end{equation}
For example, for the $E_8$ gauge group $b_0 = 90$. Taking $\alpha_{GUT}$
to have its phenomenological value given in~\eqref{check}, we obtain
\begin{equation}
\e \sim 5.
\label{GUT}
\end{equation}
The coefficients
$\alpha^{(2)}_I$ represent the tension of the hidden brane measured with respect 
to the Kahler form $\omega_{I}$~\cite{LOSW4} 
\begin{equation}
\alpha^{(2)}_I \sim  \frac{\pi \rho}{16 \pi v_{CY}} 
(\frac{\kappa_{11}}{4 \pi})^{2/3} 
 \int_{X} \omega_{I} \wedge (TrF^{(2)} \wedge F^{(2)} - \frac{1}{2} Tr R 
\wedge R) ,
\label{2.11}
\end{equation}
where $F^{(2)}$ is the curvature
of the gauge bundle on the hidden brane.
One can estimate the order of magnitude
of $\alpha^{(2)}_I$ by evaluating the right-hand-side of equation~\eqref{2.11}. 
We find 
that
\begin{equation}
\alpha^{(2)}_I \approx \frac{v_I}{v_{CY}^{1/3}},
\label{2.12}
\end{equation}
where $v_{I}$ is the volume (measured with respect to the Kahler form
$\omega_I$) of the two-cycle 
which is Poincare dual to the 
four-form $TrR \wedge R - \frac{1}{2} Tr F^{(2)} \wedge F^{(2)}$.
Similarly, the coefficient $\beta$ is the tension of the five-brane 
and given by~\cite{Visible}
\begin{equation}
\beta=\frac{2 \pi^2 \rho}{v_{CY}^{2/3}} (\frac{\kappa_{11}}{4 \pi})^{2/3}
\int_{X} \sum_{I=1}^{h^{1,1}}c_I \omega_{I} \wedge {\cal W},
\label{2.13}
\end{equation}
where ${\cal W}$ is the four-form Poincare dual to the holomorphic curve $z$ on which
the five-brane is wrapped. Evaluation of the right-hand-side of eq.~\eqref{2.13}
gives 
\begin{equation}
\beta \approx \pi \frac{v_{5}}{v_{CY}^{1/3}},
\label{2.14}
\end{equation}
where $v_{5}$ is the volume of the holomorphic curve the five-brane is 
wrapped on.
Note that $\b$ is always positive and, from~\eqref{1.8.8}, is of the same order of magnitude 
as $\tau_5$.
The real part of the combination
\begin{equation}
 S -\alpha^{(2)}_{I} T^{I} +
\b \frac{{\bf Y}^2}{{\cal T}}
\label{2.15}
\end{equation}
represents the inverse square of the gauge coupling constant on the hidden brane,
with the last two terms being the threshold corrections~\cite{LOW1, Nonstandard}.

Note that it is essential that expression~\eqref{2.15} be strictly 
positive at the vacuum of the theory. This prevents the effective gauge coupling
from diverging, or being undefined, on the hidden orbifold plane. For this 
to be the case, we must have
\begin{equation}
Re(\alpha^{(2)}_I T^I) < Re(S+ \b \frac{{\bf Y}^2}{{\cal T}}).
\label{2.15.1.1}
\end{equation}
In this paper, we want to work in the strong coupling regime of the 
heterotic string. It follows that one of the $T^I$ moduli, corresponding 
to the size of the fifth dimension, must be at least of order unity. Hence, for~\eqref{2.15.1.1}
to be satisfied, typically, we must choose the associated $\alpha^{(2)} <1$. 
We find that this can always be arranged by the appropriate choice of the vector bundle 
on the hidden orbifold plane.  
In fact, in ref.~\cite{Curiok2} it was argued that this assumption 
may be unnecessary if one includes higher order 
field-theory corrections which are protected by 
supersymmetry. This might provide 
generalizations of the results obtained in this 
paper as well.

\subsection{Non-Perturbative Superpotentials}


In this subsection, we will review the structure of non-perturbative superpotentials generated 
by strings wrapped on holomorphic curves. To be more precise, the non-perturbative contributions
to the superpotential come from membrane instantons. 
As was shown 
in~\cite{Lima1}, to preserve supersymmetry a membrane
has to be transverse to the end-of-the-world branes and
wrap a holomorphic curve in the Calabi-Yau threefold. 
In addition, only curves of genus zero 
contribute 
\cite{Witten96, DSWW2}. 
At energy scales smaller that the brane separation scale,
the membrane configuration reduces to that of a string 
wrapped on a holomorphic curve.
We will refer to such a configuration
as a heterotic string instanton. We should point out that there can be three 
different membrane configurations leading to different non-perturbative contributions to the 
superpotential. 
\begin{enumerate}
\item 
A membrane can stretch between the two orbifold fixed planes. 
\item  
A membrane can begin on the visible brane and end on the five-brane in the bulk. 
Recall that, in this paper, we are assuming that there is only one five-brane in the bulk.
\item 
A membrane can begin on the fivebrane and end on the hidden brane. 
\end{enumerate}
We will discuss the first configuration in detail and then comment
on the configurations 2. and 3. 
It was shown in~\cite{Witten00} that
the non-perturbative contribution to the superpotential of a 
string wrapped on an isolated curve $z$ has the  
structure
\begin{equation}
W_{np}[z] \propto Pfaff({\cal D}_{-}) exp(-\tau \sum_{I=1}^{h^{1,1}}
\tilde{\omega}_I T^{I}).
\label{2.16}
\end{equation}
Let us first discuss the exponential factor 
\begin{equation}
exp(-\tau \sum_{I=1}^{h^{1,1}}
\tilde{\omega}_I T^{I})
\label{2.17}
\end{equation}
which was calculated in~\cite{Lima1}.  
The coefficient $\tau$ in~\eqref{2.17} is defined as
\begin{equation}
\tau =\frac{1}{2} T_M (\pi \rho) v_z,
\label{2.18}
\end{equation}
where $T_M$ is the membrane tension given by
\begin{equation}
T_M =(2 \pi)^{1/3}(\frac{1}{2 \kappa_{11}^2})^{1/3}
\label{TM}
\end{equation}
and $v_z$ is the volume of the holomorphic curve $z$. 
By using eqs.~\eqref{1.15}, \eqref{check} and~\eqref{doublecheck}, we get
\begin{equation}
\tau \sim 250 \frac{v_z}{v_{CY}^{1/3}}.
\label{TM1}
\end{equation}
Everywhere in the paper, $\tau$ will be taken to be 
much greater than one which is naturally the case.
Furthermore, the
$\tilde{\omega}_I$ appearing in~\eqref{2.17}
are the integrals of the pullbacks to the holomorphic curve 
$z$ of the $I$-th harmonic $(1,1)$ form on Calabi-Yau threefold.
See~\cite{Lima1} for details. 
Note, that the exponential factor~\eqref{2.17}
gives the non-perturbative contribution to the superpotential 
for the $T$-moduli, but not for the Calabi-Yau volume modulus $S$. 
For example, 
when $h^{1,1}=1$, the factor~\eqref{2.17} becomes 
\begin{equation}
exp(-\tau T),
\label{2.19}
\end{equation}
where
\begin{equation}
T=R+ \frac{{\rm i}}{6}p.
\label{2.20}
\end{equation}
This shows that the superpotential associated with~\eqref{2.19}
depends on the size of the 
eleventh dimension only. 

Now let us move on to the first factor in~\eqref{2.16}. This factor, 
\begin{equation}
Pfaff({\cal D}_{-}),
\label{2.21}
\end{equation}
represents the Pfaffian of the chiral Dirac operator constructed
using the hermitian Yang-Mills connection pulled back to the curve 
$z$~\cite{Witten00, BDO1, BDO2}. 
It is clear that it depends on the vector bundle 
moduli.
So far, our discussion has been basically generic. The only restriction
on the Calabi-Yau geometry that we have made so far was to assume,
in the second half of Section 2,
that 
it is elliptically fibered. At this point,
for specificity, 
we will choose the Calabi-Yau 
threefold $X$ to be elliptically fibered over a Hirzebruch surface 
\begin{equation}
B={\mathbb F}_r.
\label{base}
\end{equation}
Let us mention some basic
properties of Hirzebruch surfaces that we will need.
The second homology group $H_2({\mathbb F}_r, {\mathbb Z})$ 
is spanned by two effective classes of curves, denoted by 
${\cal S}$ and ${\cal E}$, with intersection numbers 
\begin{equation}
{\cal S}\cdot {\cal S}= -r, \quad
{\cal S}\cdot {\cal E}= 1, \quad
{\cal E}\cdot {\cal E}= 0.
\label{intnum}
\end{equation}
The first Chern class of ${\mathbb F}_r$ is given by 
\begin{equation}
c_1({\mathbb F}_r)= 2{\cal S}+(r+2) {\cal E}.
\label{1chern}
\end{equation}
Finally, we will assume that $X$ admits a global section
$\sigma$ and that it is unique,
which is generically the case. 

A Yang-Mills vacuum consists of a stable, 
holomorphic vector bundle $V$ on the observable 
end-of-the-world brane with the structure group 
\begin{equation}
G \subseteq E_8. 
\label{E8}
\end{equation}
In general, there can be a vector bundle on the hidden brane. 
However, in this paper, we will assume that this bundle is trivial.
It follows from~\cite{Donald, UYau} that each such bundle admits 
a unique connection satisfying the hermitian Yang-Mills equations.
Over $X$ we will construct a stable, holomorphic vector bundle $V$
with structure group 
\begin{equation}
G=SU(n). 
\label{structuregroup}
\end{equation}
This is accomplished \cite{FMW, Donagi} by 
specifying a spectral cover 
\begin{equation}
{\cal C}= n \sigma + \pi^{*}\eta ,
\label{2.22}
\end{equation}
where 
\begin{equation}
\eta = (a+1){\cal S} + b{\cal E}
\label{2.23}
\end{equation}
with $a+1$ and $b$ being non-negative integers, as well as a holomorphic 
line bundle 
\begin{equation}
{\cal N} = {\cal O}_X (n (\lambda+ \frac{1}{2}) \sigma -
(\lambda -\frac{1}{2} ) \pi^* \eta +
(n\lambda +\frac{1}{2}) \pi^* c_1 ({\mathbb F}_r)),
\label{2.24}
\end{equation}
where $\lambda \in {\mathbb Z}+\frac{1}{2}$.
In eqs.~\eqref{2.22} and~\eqref{2.24}, $\pi$ is the projection map 
$\pi : X \to {\mathbb F}_r$.
Note that we use $a+1$, rather than $a$, as the coefficient of ${\cal{S}}$
in~\eqref{2.23} to conform with our conventions in~\cite{BDOold}.
We will also assume that the variables $a+1$ and $b$ satisfy the 
positivity conditions~\cite{BDOold}
\begin{equation}
a+1 >2n, \quad b> ar-n(r-2).
\label{2.25}
\end{equation}
These conditions insure that the spectral cover ${\cal C}$ 
is an ample, or positive, divisor.
The vector bundle $V$ is then determined via a Fourier-Mukai
transformation 
\begin{equation}
({\cal C}, {\cal N}) \longleftrightarrow V.
\label{2.26}
\end{equation}
The moduli of the bundle $V$ come from parameters of the spectral cover 
${\cal C}$. Since the parameters of a divisor form a complex projective space,
the moduli space of vector bundles is ${\mathbb C}{\mathbb P}^{N}$,
where $N$ is the number of the vector bundle moduli. This fact was already 
used in Section 1 in our discussion of the properties of the vector bundle 
moduli Kahler potential. 
In~\cite{BDO1, BDO2}, the Pfaffian
$Pfaff({\cal D}_{-})$
was computed in a number 
of examples for the case of a superstring wrapped on the isolated sphere 
$\sigma \cdot \pi^*{\cal S}$. 
The Pfaffian was found to be a high degree polynomial
of the vector bundle moduli. In fact, it turned out that it depends
only on a subset of the vector bundle moduli, the transition 
moduli, which are responsible for smoothing out the 
torsion free sheaf localized at the curve 
$\sigma \cdot \pi^*{\cal S}$~\cite{BDOold}.

In order to find the total non-perturbative superpotential, one has to sum 
up the contributions from all holomorphic genus zero curves, both
isolated and non-isolated.  As argued in~\cite{WitSil, BeasWit, Sethi2}, in certain cases
one can actually get zero after the summation. This makes it necessary to
discuss the genus zero holomorphic curves in Calabi-Yau 
threefolds of the type introduced above. After this discussion, we will be able 
to argue that, in these models, the superpotential does not vanish 
after the summation.
The first class of genus zero holomorphic curves are of the form
\begin{equation}
z=\sigma \cdot \pi^{*} z^{\prime},
\label{2.27}
\end{equation}
where $z^{\prime}$ is a genus zero holomorphic curve in the base
${\mathbb F}_{r}$. Below, we will often identify $z$ with $z^{\prime}$ for notation simplicity.
For specificity,
let us take the base of the Calabi-Yau threefold to be 
${\mathbb F}_{2}$. Our results will, however, remain true for other Hirzebruch surfaces as well.
The Hirzebruch surface ${\mathbb F}_{2}$, being 
a rationally ruled surface, contains one isolated genus zero curve
${\cal S}$ and infinitely many non-isolated curves. These can be shown to be  
\begin{equation}
{\cal E} {\ } {\rm and} {\ } {\cal S}+\kappa{\cal E}, 
\label{2.28}
\end{equation}
where $\kappa$ is an integer number greater than one.
Let us consider a concrete example.
In~\cite{BDO2}, it was shown that for the following choice of parameters,
\begin{equation}
n=3, \quad b-2a=3, \quad \lambda =\frac{3}{2},
\label{2.29}
\end{equation}
there are nine transition moduli, denoted by 
$\alpha_i, \beta_i, \gamma_i$,
for $i=1, 2, 3$, associated with the curve
$\sigma \cdot \pi^{*}{\cal S}$.
The Pfaffian generated by
a string wrapped on the curve 
$\sigma \cdot \pi^{*}{\cal S}$ is nonzero and given by 
the expression
\begin{equation}
Pfaff({\cal D}_{-})_{{\cal S}} = {\cal R}^{4},
\label{super}
\end{equation}
where ${\cal R}$ is the polynomial 
\begin{equation}
{\cal R}=
\alpha_1 \beta_2 \gamma_3 -\alpha_1 \beta_3 \gamma_2 +
\alpha_2 \beta_3 \gamma_1 -\alpha_2 \beta_1 \gamma_3 +
\alpha_3 \beta_1 \gamma_2 -\alpha_3 \beta_2 \gamma_1.
\label{polyn}
\end{equation}
We will now show, in the context of this example, that one can further restrict the coefficient
$a$ 
in such a way that the vector bundle moduli 
contribution to the superpotential, that is the Pfaffian, vanishes on all 
non-isolated curves of the type~\eqref{2.28}. 

As discussed in detail
in~\cite{Witten00, BDO1, Distler}, given a holomorphic genus zero curve 
$z$, the Pfaffian will vanish
if and only if the restriction of the bundle $V$ to the curve $z$ is 
non-trivial or, equivalently, that
\begin{equation}
h^{0}(z, V|_z \otimes {\cal O}_{z}(-1)) >0.
\label{2.30}
\end{equation}
It was shown in~\cite{BDO1, BDO2} that 
\begin{equation}
h^{0}(z, V|_z \otimes {\cal O}_{z}(-1)) = h^{0} (C, N(-F)|_C),
\label{2.31}
\end{equation}
where
\begin{equation}
C={\cal C}|_{\pi^{*}z}, \quad  N={\cal N}|_{\pi^{*}z}, \quad 
N(-F)= N \otimes {\cal O}_{\pi^{*}z}(-F)
\label{2.32}
\end{equation}
and $F$ is the fiber class. We will show that for non-isolated 
curves of the form~\eqref{2.28}, 
$h^{0} (C, N(-F)|_C)$
does not vanish for any value of the vector bundle moduli.
Therefore, the Pfaffian and, hence, 
the superpotential generated on such
curves will vanish identically.
The proof goes as follows. The vector space 
$H^{0}(C, N(-F)|_{C})$ lies in the exact sequence
\begin{equation}
0 \rightarrow H^{0}(\pi^{*}z, N(-F-C)) \rightarrow 
H^{0}(\pi^{*}z, N(-F)) \rightarrow H^{0} (C, N(-F)|_C) \rightarrow
\dots .
\label{2.33}
\end{equation}
It is easy to see that
\begin{equation}
h^{0}(\pi^{*}z, N(-F)) \geq h^{0}(\pi^{*}z, N(-F-C)),
\label{2.34}
\end{equation}
with the equality holding if and only if 
\begin{equation}
h^{0}(\pi^{*}z, N(-F)) =0.
\label{2.35}
\end{equation}
On the other hand, it follows from the exact sequence~\eqref{2.33}
that if 
\begin{equation}
h^{0}(\pi^{*}z, N(-F)) > h^{0}(\pi^{*}z, N(-F-C)),
\label{2.36}
\end{equation}
the dimension of the space $H^{0} (C, N(-F)|_C)$ cannot be zero and, 
therefore, the Pfaffian will vanish.
So, it is enough to show that for the curves of the form~\eqref{2.28}
the following inequality is fulfilled
\begin{equation}
h^{0}(\pi^{*}z, N(-F)) >0.
\label{2.37}
\end{equation}
Slightly abusing notation, we will 
denote the curves  
$\sigma \cdot \pi^* {\cal E}$ and 
$\sigma \cdot \pi^* ( {\cal S}+\kappa {\cal E})$ in the threefold by 
${\cal E}$ and ${\cal S}+\kappa {\cal E}$ respectively.
Using equations~\eqref{intnum}-\eqref{2.23} and~\eqref{2.29},
one can show that
\begin{equation}
N (-F)|_{\pi^{*}{\cal E}}={\cal O}_{\pi^{*}{\cal E}}
( 6\sigma|_{\pi^{*}{\cal E}}-(a-8)F)).
\label{2.38}
\end{equation}
If we demand that $a$ satisfy the positivity conditions~\eqref{2.25},
it follows from~\eqref{2.38} that condition~\eqref{2.37} 
is fulfilled for 
\begin{equation}
a=6, 7, 8.
\label{2.39}
\end{equation}
This means that, for these choices of $a$, the superpotential 
of a string wrapped on a non-isolated curve in the homology 
class of ${\cal E}$ will 
vanish for every representative in this class.
Similarly, one finds that 
\begin{equation}
N (-F)|_{\pi^{*}({\cal S}+\kappa {\cal E})}=
{\cal O}_{\pi^{*}({\cal S}+\kappa {\cal E})}
( 6\sigma|_{\pi^{*}({\cal S}+\kappa {\cal E})}
-[(a-9)\kappa +3]F).
\label{2.40}
\end{equation}
We see that condition~\eqref{2.37} is fulfilled if and only if
\begin{equation}
(a-9)\kappa +3 \leq 0, \quad \kappa >1.
\label{2.41}
\end{equation}
Equation~\eqref{2.41} is satisfied for 
\begin{equation}
a=6, 7.
\label{2.42}
\end{equation}
We conclude that, in the 
examples specified by 
\begin{equation}
n=3, \quad r=2, \quad \lambda=\frac{3}{2}, \quad b-2a=3, \quad a=6, 7,
\label{2.43}
\end{equation}
of the curves of type~\eqref{2.27}, only the isolated curve 
${\cal S}$ gives a contribution to the superpotential. 
The contributions of all non-isolated curves vanish identically
due to the vanishing of the Pfaffian. 
Even though these results have been proven within the context of a specific example, 
they are, in fact, generic, occuring for different values of $n, r, {\lambda}, a$ and $b$.

Unfortunately, the lifts of ${\cal S}, {\cal E}$ and ${\cal S}+\kappa {\cal E}$ are not the 
only genus zero, holomorphic curves in $X$. There may exist (perhaps infinitely many)
such curves contained in multisections of 
$X$\footnote{The authors are very grateful to R. Donagi and T. Pantev for 
discussions on this issue.}. These curves are regular in $X$, but project onto singular 
curves in the base. They can also be divided into two types, curves 
which are isolated in $X$ and those that are not isolated. We will denote by $\{I_x \}$
the set of such isolated curves, where $x$ indexes these curves. Similarly, 
let $\{ N_y \}$ be the set of non-isolated curves indexed by $y$. To continue our analysis,
we must consider the Pfaffian on each of these curves as well. Let us 
begin with the non-isolated curves $\{ N_y \}$. In general,
we have no reasons to believe that the Pfaffian must vanish on each of these 
curves, as it did on the non-isolated curves ${\cal E}$ and ${\cal S}+\kappa {\cal E}$
in the zero section. Therefore, these non-isolated curves may contribute
to the superpotential. However, since each such curve is non-isolated,
one must ``integrate''over the moduli of the curve. To perform such an ``integration'',
even to define it properly, is a difficult open problem. However, it has been conjectured
by Witten~\cite{Edprivatediscussions} that every non-isolated curve 
gives zero contribution to the superpotential. In this paper, we, henceforth, will assume 
that this conjecture is indeed correct and there is no 
further contribution to the superpotential arising from non-isolated 
curves $\{ N_y \}$. 

What about the isolated curves $\{I_x \}$? Generically, we expect strings wrapped around
each curve $I_x$ to produce a non-vanishing superpotential $W_{np}[I_x]$. 
The whole non-perturbative superpotential generated by 
membranes stretched between the two orbifold planes can then be written as
\begin{equation}
W_{np} =W_{np}[{\cal S}] + \sum_x W_{np}[I_x].
\label{2.44}
\end{equation}
We now want 
to make a very important 
point. For a generic Calabi-Yau threefold of the type considered here, one can show that 
none of the curves $I_x$ intersects 
${\cal S}$. That is, 
\begin{equation}
{\cal S} \cdot I_x =0
\label{horline}
\end{equation}
for all values of $x$.
This leads to the following conclusion. That is, the 
superpotentials $W_{np}[{\cal S}]$ and $\sum_x W_{np}[I_x]$ depend on 
different vector bundle moduli. Let $\phi_i$ be the transition moduli
associated with the curve ${\cal S}$. Since ${\cal S}$ and all $I_x$
do not intersect, they do not share transition moduli. Therefore,
the sum $\sum_x W_{np}[I_x]$ does not depend on $\phi_i$. Similarly,
$W_{np}[{\cal S}]$ does not depend on the vector bundle moduli associated with any of the curves 
$I_x$. Let us now split the vector bundle moduli $\Phi_u$ as
\begin{equation}
\Phi_u =\{\phi_i, \chi_{\tilde{i}}, \psi_a\},
\label{2.45}
\end{equation}
where $\phi_i$ are the transition moduli associated with the curve 
${\cal S}$, $\chi_{\tilde{i}}$ are the transition moduli associated with
all the curves $I_x$ and $\psi_a$ are the remaining moduli. Since we do not 
expect that the whole second Chern class of the bundle $V$ is 
localized on the isolated curves, the moduli $\phi_i$ and 
$\chi_{\tilde{i}}$ do not span the entire moduli space.
We can now rewrite the superpotential~\eqref{2.44} as
\begin{equation}
W_{np} = W_{np}(\phi) +W_{np}(\chi),
\label{2.46}
\end{equation}
where the vector bundle moduli $\phi_i$ and $\chi_{\tilde{i}}$ do not overlap.
Note that $W_{np}$ is independent of the moduli $\psi_a$.
This, in particular, shows that the non-perturbative superpotential
is not zero if at least one of the terms is not zero.  
The term $W_{np}(\phi)$ was calculated and found to be 
non-vanishing in a number of examples in~\cite{BDO1, BDO2}. 
For example, the Pfaffian on ${\cal S}$ was computed to be~\eqref{super}
in our $B={\mathbb F}_2$ example above.
This means that, 
in such examples, the non-perturbative superpotential is not zero provided 
the conjecture about the vanishing of the superpotential on 
non-isolated curves is indeed correct. 

Let us now give the generalization of the above discussion to the case when a
membrane stretches between one of the orbifold planes and a five-brane. 
As we have said, in this paper we will assume that there is a single five-brane in the bulk.
The non-perturbative superpotential for such a membrane configuration was calculated
in~\cite{Lima2, Moore}. The contribution has a 
form very similar to~\eqref{2.16}. When a membrane begins on the observable brane 
and ends on the five-brane wrapped on an isolated genus zero 
holomorphic curve $z$, the superpotential is
\begin{equation}
W^{(1)}_5 \propto Pfaff({\cal D}_{-})e^{-\tau {\bf Y}},
\label{2.47}
\end{equation}
where ${\cal D}_{-}$ is the chiral Dirac operator associated with the bundle $V$ 
on the observable brane restricted to $z$ and the coefficient $\tau$ is given in~\eqref{2.18}.
When the membrane stretches between the five-brane
and the hidden brane, the superpotential will be
\begin{equation}
W^{(2)}_5 \propto Pfaff({\cal D}^{hidden}_{-})e^{-\tau ({\cal T}-{\bf Y})}.
\label{2.48}
\end{equation}
By $Pfaff({\cal D}^{hidden}_{-})$, we denote the Pfaffian of the Dirac
operator constructed using the pullback to $z$ of the hermitian Yang-Mills 
connection on the hidden brane. If the vector bundle on the hidden brane 
brane is trivial, as we are assuming in this paper, the Pfaffian is simply a 
constant, independent of moduli, and the corresponding 
contribution to the superpotential becomes 
\begin{equation}
W^{(2)}_5 \propto e^{-\tau ({\cal T}-{\bf Y})}.
\label{2.49}
\end{equation}

Before closing this section, we want to be a little more
explicit about what we mean by the assumption that there is a single five-brane in the 
bulk. For a trivial vector bundle on the hidden brane, the anomaly cancellation
condition determines the five-brane class to be 
\begin{equation}
{\cal W} = c_2(TX) -c_2(V).
\label{anomaly}
\end{equation}
%
It was shown in~\cite{Five} that the moduli space of the homology class
${\cal W}$ always contains an irreducible representative curve. Physically, this 
corresponds to a single five-brane in the bulk space. In this paper, 
we always take the five-brane to be wrapped on an irreducible curve. 


\section{Moduli Stabilization}


\subsection{Setting Up a Model}


In this section, we will provide a stabilization of the moduli considered above.
Unfortunately, the Kahler potentials~\eqref{1.7} and~\eqref{1.8.5} and the
superpotentials~\eqref{2.16}, \eqref{2.46}, \eqref{2.47} and~\eqref{2.49}
are very complicated when the number of the $(1,1)$ moduli $T^I$ 
and the vector bundle moduli $(\phi_i, \chi_{\tilde{i}})$ is large. For the 
case of Calabi-Yau threefolds elliptically fibered over the Hirzebruch surfaces,
the number of the $T$-moduli is three and the number of the vector bundle moduli is 
of order one hundred or larger~\cite{BDOold}. The number of transition moduli associated with 
the curve ${\cal S}$ is also quite large, of order ten~\cite{BDOold}.
Therefore, to give an explicit analytic solution, we have to 
simplify the model without losing its essential properties. Our first step in this direction will be to 
assume that we have only one $(1,1)$ modulus.
We can do this without loss of generality since, as will become clear in our analysis, any number of the 
$T$-moduli can be stabilized by
the same mechanism. Let us emphasize that
the reason for doing this is purely technical. We just want to simplify the equations.
We will comment on this further at the appropriate place. Henceforth, we will
take only one $T$-modulus, which is associated with the size of the eleventh dimension.
We now have to make some simplifications concerning the vector bundle moduli.
In the previous section, we split the vector bundle moduli $\Phi_u$ 
into three categories, the transition moduli, $\phi_i$, associated with the curve ${\cal S}$,
the transition moduli, $\chi_{\tilde{i}}$, associated with the curves $\{ I_x \}$
and the remaining moduli $\psi_a$. Clearly, the equations of motion for the 
$\phi_i$- and $\chi_{\tilde{i}}$- moduli are very similar. 
Therefore, we may assume that there are no 
$\chi_{\tilde{I}}$-moduli at all without any loss of generality. If we manage to stabilize 
the $\phi_i$-moduli, the moduli $\chi_{\tilde{i}}$ will be stabilized by precisely the same procedure. 
Ignoring the $\chi$-moduli does not produce conceptual changes in the structure of the 
$T$- and $Y$-superpotentials either. Specifically, the existence of the $\chi$ terms in~\eqref{2.46},
in addition to the first term, can at most produce a racetrack potential energy 
for the $T$- and $Y$-moduli. This would only strengthen the vacuum stability.
However, we must continue to keep the $\psi_a$-moduli, 
since they do not appear in any of the superpotentials discussed in the previous section
and ignoring them can conceptually alter the potential energy. 
As a result, we will assume that $\phi_i$ and $\psi_a$ are
all of the vector bundle moduli. We now want to 
introduce simplifications concerning the number of $\phi_i$-moduli. In all of the 
examples studied in~\cite{BDO1, BDO2}, it was found that the number of  
transition moduli associated with the curve ${\cal S}$ is large
and that the corresponding Pfaffian is a complicated homogeneous polynomial 
of high degree. Again, for simplicity, we will pretend that there is only one 
$\phi$-modulus. From the discussion of its stabilization, it will be obvious that any 
number of $\phi$-moduli can be stabilized by the same mechanism. 
Therefore, we can restrict to a single $\phi$ modulus without any loss of generality.
To conclude, we consider a model
containing the following moduli. We have one $T$-modulus geometrically 
corresponding to the separation of the orbifold planes, the $S$-modulus corresponding 
to the Calabi-Yau volume, $h^{1,2}$ moduli $Z_{\alpha}$ whose precise number is irrelevant,
one transition vector bundle modulus $\phi$, the remaining vector bundle moduli
$\psi_a$ whose precise number is also irrelevant and one five-brane modulus ${\bf Y}$.
We emphasize, once again, that these simplifications are made for purely technical reasons
to simplify the equations. Any number of the $T$-, $\phi$- and $\chi$-moduli
can be stabilized by a similar method. 

Let us write the simplified Kahler potential and the superpotential relevant for our model.
We have 
\begin{equation}
K=K_{S,T}+K_Z+K_5+ k K_{bundle}.
\label{3.1} 
\end{equation}
In this expression, $K_{S,T}$ is given by (see eq.~\eqref{1.7})
\begin{equation}
K_{S,T}= -M^{2}_{Pl} \ln (S+\bar S) - 
3 M^{2}_{Pl} (T +\bar T),
\label{3.2}
\end{equation}
where eq.~\eqref{1.4} has been used. 
In order to ignore the cross term~\eqref{cross5}, we must always work in a region of moduli space where
\begin{equation}
|Im T| \ll 1.
\label{im}
\end{equation}
The $h^{2,1}$ moduli Kahler potential $K_Z$ is given 
in~\eqref{1.8} by 
\begin{equation}
K_{Z}= -M^{2}_{Pl} \ln (-i \int_{X} \Omega \wedge \bar \Omega ).
\label{3.3}
\end{equation}
The five-brane Kahler potential $K_5$~\eqref{1.8.5} now becomes
\begin{equation}
K_5 = 2 M^{2}_{Pl} \tau_5 \frac{({\bf Y} +\bar{\bf Y})^2}{(S+\bar S)(T+\bar T)}.
\label{3.4}
\end{equation}
By definition (see eq.~\eqref{1.8.1}) 
\begin{equation}
0\leq Re{\bf Y}\leq Re T,
\label{re}
\end{equation}
since the five-brane must be between the orbifold planes.
The vector bundle moduli Kahler potential is not known explicitely. However, from 
our discussion at the end of Section 2, we concluded that $K_{bundle}$
can be split as follows
\begin{equation}
K_{bundle} =K_{bundle} (\phi) + K_{bundle} (\psi_a).
\label{3.5}
\end{equation}
We also know that for small values of $\phi$, $K_{bundle}$ must diverge. 
For concreteness, 
when $\phi$ is sufficiently less than one, we take $K_{bundle}(\phi)$ to be 
\begin{equation}
K_{bundle}(\phi) =-p \ln (\phi + \bar \phi),
\label{3.6}
\end{equation}
where $p$ being some dimensionless, positive constant. Expression~\eqref{3.6} 
is the function that diverges most softly at zero. However, one can 
choose $K_{bundle}(\phi)$ to be any other function that diverges at zero, 
for example, an inverse polynomial in $\phi$. 
We can show that $\phi$ can be stabilized for any such functions. 

Let us now summarize the superpotential. The total superpotential
is given by 
\begin{equation}
W =W_f + W_g + W_{np} +W^{(1)}_5 +W^{(2)}_5.
\label{3.7}
\end{equation}
Here, $W_f$ is the flux-induced superpotential (see eq.~\eqref{2.6})
\begin{equation}
W_{f} = 
M^{3}_{Pl} h_1 \int_{X} \tilde{H} \wedge 
\tilde{\Omega},
\label{3.8}
\end{equation}
where 
\begin{equation}
h_1 \sim 2 \cdot 10^{-8}.
\label{h1}
\end{equation}
$W_g$ is the superpotential induced by the gaugino condensation
on the hidden wall. In our model, it follows from~\eqref{2.8} that
\begin{equation}
W_g =M_{pl}^3 h_2 exp (-\e S +\e \alpha^{(2)}T -
\e \b \frac{{\bf Y}^2}{T}),
\label{3.9}
\end{equation}
where (see eqs.~\eqref{2.9}, \eqref{2.12} and~\eqref{2.14})
\begin{equation}
h_2 \sim 10^{-6}, \quad \alpha^{(2)} \sim \frac{v}{v_{CY}^{1/3}}, \quad 
\b \sim \pi \frac{v_5}{v_{CY}^{1/3}}.
\label{3.10}
\end{equation}
According to our discussion in the previous section, to make sure that the combination
\begin{equation}
Re(S-\alpha^{(2)}T +\b \frac{{\bf Y}^2}{T})
\label{3.10.1}
\end{equation}
is positive, we have to take $\alpha^{(2)}$ to be less than one. 
The non-perturbative superpotential $W_{np}$ (eq.~\eqref{2.46}) is now given by
\begin{equation}
W_{np} =c_1 M_{Pl}^3 \phi^{d+1} e^{-\tau T},
\label{3.11}
\end{equation}
where we have restored its natural scale and $c_1$ is some dimensionless coefficient of order unity.
The Pfaffian, which must be a homogeneous polynomial, is represented by the 
factor $\phi^{d+1}$. We will assume 
that $d+1$ is sufficiently large. 
This is naturally the case in explicit examples~\cite{BDO1, BDO2}.

To discuss the five-brane superpotentials, we must first specify the holomorphic curve over which 
the five-brane is wrapped. As emphasized above, that curve can always be chosen to be irreducible 
corresponding to a single five-brane. However, in general the homology class 
${\cal W}$ of the curve can contain both horizontal and components,
involving ${\cal S}, {\cal E}$ and the fiber $F$ respectively. We find it easiest 
to chose ${\cal W}$ to be simply at least one copy of the curve ${\cal S}$. This 
can always be accomplished by adjusting the bundle $V$ on the observable brane. 
Henceforth, in this paper, we will assume that this is the case. The more general 
case is more difficult to analyze and will be presented elsewhere.
The five-brane non-perturbative superpotentials $W_5^{(1)}$ and $W_5^{(2)}$ 
(eqs.~\eqref{2.47}-\eqref{2.49}) are then given by 
\begin{equation}
W_5^{(1)} = c_2 M^3_{Pl} \phi^{d+1} e^{-\tau {\bf Y}}
\label{3.12}
\end{equation}
and 
\begin{equation}
W_5^{(2)} = c_3 M^3_{Pl} e^{-\tau (T-{\bf Y})},
\label{3.13}
\end{equation}
with $c_2$ and $c_3$ being dimensionless coefficients of order unity.
In eq.~\eqref{3.13}, we have assumed that the bundle is trivial on the hidden brane.
Note that the Pfaffian in $W_5^{(1)}$ is identical to the one in~\eqref{3.11} since both
arise from the Dirac operator restricted to the curve ${\cal S}$.


\subsection{Moduli Stabilization}


In this section, we will show that the system of the equations
\begin{equation}
D_{all{\ }fields}W=0,
\label{3.14}
\end{equation}
where $DW$ is the Kahler covariant derivative
\begin{equation}
DW=\partial W +\frac{1}{M^2_{Pl}} (\partial K)W,
\label{3.15}
\end{equation}
has a solution in the correct phenomenological range for all fields. In other words, 
we will show that all moduli described earlier can be stabilized 
in an AdS vacuum. 

We start with the system of equations
\begin{equation}
D_{\psi_a}W=0.
\label{3.16}
\end{equation}
Since the superpotential $W$ does not depend on $\psi_a$, the above equations are reduced
to 
\begin{equation}
\frac{\partial K_{bundle}(\psi)}{\partial \psi_a} =0,
\label{3.17}
\end{equation}
where eq.~\eqref{3.5} has been used.
We will now argue that this equation always has a solution.
From Section 2, 
we know that as $\psi$ goes to positive infinity 
along either its real or imaginary directions,
the Kahler potential $K_{bundle}$ grows. 
On the other hand, as 
$\psi$ goes to zero, $K_{bundle}$ can either stay regular 
or diverge. 
$K_{bundle}$ will diverge if
the locus $\psi =0$ corresponds to a torsion free
sheaf supported on a 
holomorphic curve different from that 
associated with the vanishing of $\phi$. 
If $K_{bundle}$ diverges at zero, 
then equation~\eqref{3.17} must have a solution 
for positive $\psi$ 
corresponding to a 
minimum of the function $K_{bundle}$. 
If $K_{bundle}$ is a regular function 
of $\psi$ at zero, we can ask 
what happens as $\psi$ grows in its negative
real or imaginary directions. From equations~\eqref{1.24},~\eqref{1.25}
and the properties of the partition of unity, it follows that 
$K_{bundle}$ must also grow
in these negative directions. Therefore, again, $K_{bundle}$ 
must have a minimum. 
Thus the properties of ${\mathbb C}{\mathbb P}^{N}$ 
guarantee the existence 
of a solution to equation~\eqref{3.17}. 

As the second step, consider the equations involving the 
Kahler covariant derivative with respect to the complex structure moduli 
$Z_{\alpha}$
\begin{equation}
D_{Z_{\alpha}}W=0,
\label{3.18}
\end{equation}
which is equivalent to 
\begin{equation}
\partial_{Z_{\alpha}}W_f +\frac{1}{M^2_{Pl}} (\partial_{Z_{\alpha}} K_Z)W=0.
\label{3.19}
\end{equation}
Let us make an assumption that the absolute value of $W_f$ is sufficiently larger than 
all other contributions to the superpotential near the vacuum, that is
\begin{equation}
|W_f| \gg |W_g|, |W_{np}|, |W_5^{(1)}|, |W_5^{(2)}|.
\label{A}
\end{equation}
Later, we will see that our solution is completely consistent with this assumption. 
Then eq.~\eqref{3.18} becomes
\begin{equation}
\partial_{Z_{\alpha}}W_f +\frac{1}{M^2_{Pl}} (\partial_{Z_{\alpha}} K_Z)W_f=0.
\label{3.20}
\end{equation}
All terms in this equation depend only on the complex structure moduli. It is 
not known how to find either $K_Z$ or $W_f$ for complicated 
Calabi-Yau geometries.  Both $K_Z$ and $W_f$ are expected to be
complicated functions of $Z_{\alpha}$.
Nevertheless, there is evidence that 
this system of equations has a non-trivial solution. First, the number of equations 
is equal to the number of the unknowns, so one can expect a solution. A second piece of evidence 
comes from ref.~\cite{Schulz}, where type IIB flux compactifications 
on the space of rather simple geometry $T^6/Z_2$
were considered. In that paper, it was shown that an equation analogous to~\eqref{3.20}
indeed has a solution fixing all the moduli $Z_{\alpha}$. Thus, we simply assume that  
eqs.~\eqref{3.20} fix all the complex structure moduli $Z_{\alpha}$ and that the 
value of $W_f$ at the minimum is nonzero. Exactly the same assumption was crucial
for the moduli stabilization in the type IIB theory discussed in~\cite{Kachru}.

Before moving on to the other equations, let us introduce some notation. Let 
\begin{equation}
T = T_1 +i T_2, \quad S =S_1 +i S_2, \quad {\bf Y}={\bf Y_1} +{\bf Y_2}, \quad
\phi =re^{i \theta}.
\label{not1}
\end{equation}
Also, write the value of $W_f$ in the minimum as 
\begin{equation}
W_f = |W_f|e^{i f},
\label{not2}
\end{equation}
that is, we write the complex number $W_f$ in terms of its absolute value and its phase.
Now consider the equation
\begin{equation}
D_S W=0.
\label{3.21}
\end{equation}
By using eqs.~\eqref{3.2} and~\eqref{3.9}, we obtain
\begin{equation}
(2\e S_1)W_{g} =- W_f.
\label{3.22}
\end{equation}
This complex equation is equivalent to two real equations, one relating the 
phases of the left- and right-hand-sides, the other relating the absolute values.
The phase equation is as follows
\begin{equation}
-\e S_2 +\e \alpha^{(2)} T_2 -\e \b Im (\frac{{\bf Y}^2}{T})=f +\pi(2n_1+1),
\label{3.23}
\end{equation}
where $n_1$ is any integer. Here we have used the notation introduced in~\eqref{not1} 
and~\eqref{not2}. The absolute value equation, on the other hand, is 
\begin{equation}
(2 \e S_1)|W_g|=|W_f|.
\label{3.24}
\end{equation}
%
From this equation we obtain a solution for $S_1$ as a function of $T$ and ${\bf Y}$. 
Taking, for simplicity, 
\begin{equation}
Re(\alpha^{(2)} T -\b \frac{{\bf Y}^2}{T}) \ll S_1
\label{3.24.1}
\end{equation}
we find 
\begin{equation}
(2 \e S_1) e^{-\e S_1} =\frac{|W_f|}{h_2}.
\label{3.25}
\end{equation}
This equation provides a solution for $S_{1}$. This mechanism is similar to that considered 
in~\cite{Kachru}. The scale that controls $W_f$ 
was found in Section 3 to be $h_1 \sim 2 \cdot 10^{-8}$, whereas $h_2$ was found to be 
$h_2 \sim 10^{-6}$. Therefore, we find that
\begin{equation}
(\e S_1) e^{-\e S_1} \sim \frac{h_1}{h_2} \sim 10^{-2}.
\label{3.26}
\end{equation}
From here we obtain
\begin{equation}
\e S_1 \approx 7.5.
\label{3.27}
\end{equation}
Recalling from~\eqref{GUT}
that for the $E_8$ gauge group $\e$ is of order $5$, it follows that 
\begin{equation}
S_1 \approx 1.5,
\label{3.27.1} 
\end{equation}
a phenomenologically accepted solution for $S_1$. 
We also see from~\eqref{3.25} that, by turning on a larger 
flux, we can reduce the value of $S_1$ to make it closer to one. 
It is clear that one can find a solution for $S_1$ of order unity for generic 
values of 
\begin{equation}
Re(\alpha^{(2)} T -\b \frac{{\bf Y}^2}{T})
\label{3.27.2}
\end{equation}
less than $S_1$.
We should point out that, in principle,
the absolute value of the flux superpotential in the minimum can be less, and even much less 
(in units of $M_{Pl}^3$), than the order of $h_1$. Then there are 
two possibilities. First, we can turn on a larger amount of the flux,
thus increasing $|W_f|$,
to keep $\e S_1$ at the 
same value as in~\eqref{3.27}, that is, of order ten. The other possibility is that we can put a non-trivial
bundle on the hidden brane. It will break the low-energy gauge group 
on the hidden wall from $E_8$ down to some proper subgroup. All this will reduce $b_0$ 
and, therefore, from~\eqref{2.10} we see that this will increase $\e$. Hence, we 
still can solve eq.~\eqref{3.25} and find $\e S_1 \sim 10$. Note that, since
$\e S_1$ is of order ten, we see from~\eqref{3.24} that
$|W_f|\gg |W_g|$. This is in agreement with the relevant part of assumption~\eqref{A} which
we made to justify~\eqref{3.20}. 

Now let us consider the equation
\begin{equation}
D_{\phi} W=0.
\label{3.28}
\end{equation}
By using eqs.~\eqref{3.5}, \eqref{3.6}, \eqref{3.11}, \eqref{3.12} and \eqref{A},
we obtain the following expression
\begin{equation}
(d+1) \phi^{d}(c_1 e^{-\tau T} +c_2 e^{-\tau {\bf Y}}) = \frac{pk}{\phi + \bar \phi} W_f.
\label{3.29}
\end{equation}
Since $\tau$ given in~\eqref{2.18} is always much larger than one, the first term on the 
left-hand-side of~\eqref{3.29}
is much smaller than the second as long as $\bf Y_{1} \rm <T_{1}$. We will assume here that 
this is the case, justifying this assumption later on.
Then, approximately, we have 
\begin{equation}
W_5^{(1)} e^{-i \theta} = \frac{pk}{2(d+1)\cos \theta} W_f,
\label{3.30}
\end{equation}
where we have used~\eqref{3.12} and~\eqref{not1}.
As before, this complex equation is equivalent to two real 
equations, one relating the phases of $W_5^1$ and $W_f$ and one relating their 
absolute values. The phase equation reads
\begin{equation}
d\theta -\tau {\bf Y_2} = f +2 \pi n_2,
\label{3.31}
\end{equation}
where $n_2$ is any integer. The equation for the absolute value is 
\begin{equation}
|W_5^{(1)}|=\frac{pk}{2(d+1) \cos \theta} |W_f|.
\label{3.32}
\end{equation}
Eqs.~\eqref{3.31} and~\eqref{3.32} stabilize the vector bundle moduli 
$r$ and $\theta$ provided the five-brane moduli ${\bf Y_1}$ and ${\bf Y_2}$
are stabilized. Note that, since $k \sim 10^{-5}$ and $(d+1)$ is large, we 
have $|W_5^{(1)}| \ll |W_f|$ for generic values of $c_2$ and $\cos \theta$.
Similarly, since the first term in~\eqref{3.29}
is proportional to $W_{np}$, it follows that
$|W_{np}| \ll |W_f|$.
This is consistent with our assumption~\eqref{A}. 
At this point, we would like to discuss what would happen if we took an arbitrary 
number, say $M$, of $\phi$-moduli. Eqs.~\eqref{3.31} and~\eqref{3.32} would be
two sets of $M$ equations, one for $M$ phases $\theta_i$ and one for $M$ radii $r_i$. 
For the phases, we would have $M$ equations of the type~\eqref{3.31} which 
would determine all $\theta_i$'s as functions of ${\bf Y_2}$. 
Similarly, we would have $M$ inhomogeneous equations for $r_i$ of the type~\eqref{3.32}.
Clearly, for a generic Pfaffian one expects to find a solution.
Furthermore, a generic Kahler potential $K_{bundle}(\phi)$ would not drastically modify  
eq.~\eqref{3.32}. It would still be an inhomogeneous 
equation for $r$ (or $r_i$'s in case there are several)
and one still expects a solution. It is also clear that the omitted 
moduli $\chi_{\tilde{i}}$ can be stabilized by the same mechanism.

Let us move on to the equation
\begin{equation}
D_{T}W=0.
\label{3.33}
\end{equation}
By using eqs.~\eqref{3.2}, \eqref{3.4}, \eqref{3.11}-\eqref{3.13}, \eqref{A}
and the fact that $(\frac{{\bf Y_1}}{T_1})^2$ is sufficiently less than one, we obtain 
\begin{equation}
2 \tau T_1 W_{5}^{(2)}= -3 W_f.
\label{3.34}
\end{equation}
This equation is very similar to eq.~\eqref{3.22}. Relating the phases
of the left- and right-hand-sides of~\eqref{3.34} gives
\begin{equation}
-\tau T_2 +\tau {\bf Y_2} = f +\pi (2n_3 +1),
\label{3.35}
\end{equation}
where $n_3$ is any integer. Relating the absolute value yields
\begin{equation}
2 \tau T_1 |W_{5}^{(2)}|= 3 |W_f|.
\label{3.36}
\end{equation}
or, more precisely,
\begin{equation}
2 c_3 \tau T_1 e^{-\tau (T_1-{\bf Y_1})} = 3 W_f.
\label{3.37}
\end{equation}
Since we take $\tau$ to be much greater than one and 
$|W_f|$ is much less than one, we can always find a solution for $T_1$
in the correct phenomenological range of order one by adjusting the 
parameter $\tau$, provided
${\bf Y_1}$ is stabilized. It is clear that a similar consideration would hold 
in the case of several $T$-moduli, though the equations would be more 
complicated. Note that, since $\tau \gg 1$, it follows that $|W_f| \gg |W_5^{(1)}|$. Therefore,
all conditions in assumption~\eqref{A} are satisfied. 

The last equation to consider is 
\begin{equation}
D_{{\bf Y}}W =0.
\label{3.38}
\end{equation}
By using eqs.~\eqref{3.4}, \eqref{3.9}, \eqref{3.12}, \eqref{3.13}, \eqref{A}, \eqref{3.22},
\eqref{3.24}, \eqref{3.34} and~\eqref{3.36}, we get the following equation
\begin{equation}
\frac{\b {\bf Y}}{S_1 T} -\frac{3}{2 T_1} + 2 \tau_5 \frac{{\bf Y_1}}{T_1S_1}=0.
\label{3.39}
\end{equation}
Since, to justify dropping the cross term~\eqref{cross5},
we are looking for a solution with $|T_2| \ll 1$, we have
\begin{equation}
\frac{{\bf Y}}{T} \approx \frac{{\bf Y_1} +i {\bf Y_2}}{T_1}
\label{3.40}
\end{equation}
Then the imaginary part of eq.~\eqref{3.39}
becomes
\begin{equation}
{\bf Y_2} \approx 0.
\label{3.41}
\end{equation}
This provides the stabilization of ${\bf Y_2}$.
The real part of eq.~\eqref{3.39} reads
\begin{equation}
\frac{\b {\bf Y_1}}{S_1} -\frac{3}{2} + 2 \tau_5 \frac{{\bf Y_1}}{S_1} =0.
\label{3.42}
\end{equation}
From here we get
\begin{equation}
{\bf Y_1} =\frac{3S_1}{2\b  + 4 \tau_5}.
\label{3.43}
\end{equation}
This is the solution for ${\bf Y_1}$, provided it satisfies
\begin{equation}
{\bf Y_1}< T_1.
\label{3.44}
\end{equation}
to justify our previous assumption.
Taking, for example, $T_1$ of order one and noticing that
both $\b$ and $\tau_5$ are of the same order of magnitude
(from eqs.~\eqref{1.8.8} and~\eqref{2.14} we see that they are both of order 
$\frac{v_5}{v_{CY}^{1/3}}$), \eqref{3.44} leads to the following condition on $\beta$
\begin{equation}
\beta > \frac{S_1}{2}.
\label{3.45}
\end{equation}
This is the condition on the coefficient $\beta$ in order to make sure that 
${\bf Y_1}$ is stabilized in the correct range. Taking, for example,
\begin{equation}
S_1 \sim 1, \quad T_1 \sim 1,
\label{3.46}
\end{equation}
we obtain
\begin{equation}
\b > 0.5,
\label{3.47}
\end{equation}
which is a rather mild condition since $\b$ is generically of order one.
If the condition~\eqref{3.45} is not satisfied then, at least in the 
low-energy field theory approximation,
the five-brane is pushed all the way to the hidden brane. 
Let us make sure that we have indeed stabilized the absolute value of the 
imaginary part $T_2$ at a value much less than one.  
From eqs.~\eqref{3.35} and~\eqref{3.41}, we get
\begin{equation}
T_2 =- \frac{f +\pi(2 n_3 +1)}{\tau}.
\label{3.48}
\end{equation}
Since $\tau$ is much greater than one, we can use our freedom to adjust the integer 
$n_3$ to make $|T_2| \ll 1$, which justifies dropping the cross term~\eqref{cross5} in the Kahler potential. 
Since the imaginary
part of the five-brane modulus ${\bf Y_2}$ was found in~\eqref{3.41} to be 
approximately zero, we can write the 
solution~\eqref{3.31} and~\eqref{3.32} for the phase $\theta$ and the
absolute value $r$ as follows
\begin{equation}
\theta =\frac{f + 2\pi n_2}{d}
\label{3.49}
\end{equation}
and 
\begin{equation}
r =\left ( \frac{pk|W_f|e^{\tau {\bf Y_1}}}{2(d+1)c_2 \cos \theta } \right )^{1/d}.
\label{3.50}
\end{equation}
Similarly, the imaginary part of the $S$-modulus can be easily found from eq.~\eqref{3.23} to be
\begin{equation}
S_2 \sim -\frac{f+2\pi n_1}{\e},
\label{3.51}
\end{equation}
where we have used eq.~\eqref{3.41} and the fact that $|T_2|$ is much less than one. 
Thus, when~\eqref{3.45} is satisfied,
we have found a stable solution for all of the heterotic M-theory moduli. 

Let us summarize our solution. In this section, we found a stable AdS
minimum for all heterotic M-theory moduli, namely the complex structure 
moduli $Z_{\alpha}$, the dilaton $S$,
the $h^{1,1}$ modulus $T$, the vector bundle moduli $\phi$ and $\psi_a$ and the five-brane modulus 
${\bf Y}$. The complex structure moduli are fixed by the fluxes. The corresponding equations have the 
standard form~\eqref{3.20}. 
The real part of the dilaton, $S_1$, is obtained by solving eq.~\eqref{3.25}. 
As explained below eq.~\eqref{3.25},
one can stabilized $S_1$ near its phenomenological value of order one. 
The imaginary part of $S$ is given by eq.~\eqref{3.51}.
The 
real part of the $T$-modulus is stabilized in a 
similar way by solving eq.~\eqref{3.37} together with the eq.~\eqref{3.43} for the five-brane modulus
${\bf Y_1}$. Clearly, one can stabilize $T_1$ at a value near its phenomenological value of order one.
For example, if we take 
\begin{equation}
c_3 \approx 1, \quad S \approx 1, \quad \b \approx 0.8,  
\label{example1}
\end{equation}
from eqs.~\eqref{1.8.8}, \eqref{2.14}, \eqref{TM1}, \eqref{3.37} and~\eqref{3.43}
we find
\begin{equation}
T_1 \approx 0.7, \quad {\bf Y_1} \approx 0.6.
\label{example2}
\end{equation}
The imaginary parts of both the $T$-modulus and the ${\bf Y}$-modulus are stabilized at values close to zero.
The phase and the absolute value of the vector bundle modulus $\phi$ are given in
eqs.~\eqref{3.49} and~\eqref{3.50} respectively.
The vector bundle moduli $\psi_a$ are stabilized by the properties of ${\mathbb C}{\mathbb P}^N$,
as explained below eq.~\eqref{3.17}. Remarkably,
the only constraint that we have to impose on the various coefficients is given in eq.~\eqref{3.45}, 
which is easily satisfied.

Finally, it is straightforward to write the value of the potential energy at the minimum.
It is given by the equation
\begin{equation}
V_{min} = -3 e^{K/M^{2}_{Pl}} \frac{|W|^2}{M^2_{Pl}} \sim -\frac{|W_f|^2}{M^2_{Pl}},
\label{3.52}
\end{equation}
where eq.~\eqref{A} has been used. The size of the potential energy is determined
by the value of the flux-induced superpotential. 
Since the scale that controls $W_f$ is of order $10^{-8}$ (in units of $M_{Pl}^3$), we expect 
$V_{min}$ to be 
\begin{equation}
V_{min}  \sim  - 10^{-16} M^{4}_{Pl} \sim -10^{60} (GeV)^4.
\label{3.53}
\end{equation}
Clearly, the masses of the excitations
around this minimum are also determined by the fluxes.


\section{Conclusion}


In this paper, we have shown that all moduli of strongly coupled heterotic string theory 
can be stabilized with vacuum expectation values in a phenomenologically accepted range. 
This vacuum preserves ${\cal N}=1$ supersymmetry in the moduli sector, but has a rather 
deep negative cosmological constant whose scale is set by the compactification mass. 
Supersymmetry is, however, softly broken in the gravity and matter sectors at the TeV 
scale by the gaugino condensate.
Our result is the heterotic string analog of the supersymmetry preserving part 
of the stabilization procedure presented in the type IIB context in~\cite{Kachru}.
There are, however, a number of new, non-trivial elements in the heterotic discussion. 
These include the vector bundle moduli and their non-perturbative superpotentials, the 
gaugino condensate superpotential with threshold corrections and the inclusion of a bulk five-brane
and its non-perturbative dynamics. 

It is natural to ask whether, by appropriate modification of our heterotic theory, the 
value of the potential energy at the local minimum can be lifted from its large, 
negative value to a small, positive cosmological constant of the order, say, of dark energy.
This was accomplished in the type IIB context in~\cite{Kachru} by adding anti D-branes.
It would be interesting to try to find a heterotic analogue of this mechanism involving 
anti M five-branes. Alternatively, one could try to use the mechanisms recently proposed in~\cite{Bur}
to lift the vacuum to a positive value. We will discuss this elsewhere.


\section{Acknowledgements}


The authors would like to thank V. Balasubramanian, R. Donagi, A. Lukas, J. Maldacena, T. Pantev,
R. Reinbacher, D. Waldram and E. Witten for lots of helpful discussions  
and G. Curio and A. Krause for e-mail correspondence.
The work of E.~I.~B. is supported by NSF grant PHY-0070928. The work of B.~A.~O. is supported 
in part by the DOE under contract No. DE-AC02-76-ER-03071 and by the 
NSF Focussed Research grant DMS0139799. 



\end{document}